\documentclass[twoside,twocolumn,9pt]{article}
\usepackage{extsizes}
\usepackage[super,sort&compress,comma]{natbib} 
\usepackage[version=3]{mhchem}
\usepackage[left=1.5cm, right=1.5cm, top=1.785cm, bottom=2.0cm]{geometry}
\usepackage{balance}
\usepackage{mathptmx}
\usepackage{sectsty}
\usepackage{graphicx} 
\usepackage{lastpage}
\usepackage[format=plain,justification=justified,singlelinecheck=false,font={stretch=1.125,small,sf},labelfont=bf,labelsep=space]{caption}
\usepackage{float}
\usepackage{fancyhdr}
\usepackage{fnpos}
\usepackage[english]{babel}
\addto{\captionsenglish}{%
  
}
\usepackage{array}
\usepackage{droidsans}
\usepackage{charter}
\usepackage[T1]{fontenc}
\usepackage[usenames,dvipsnames]{xcolor}
\usepackage{setspace}
\usepackage[compact]{titlesec}
\usepackage{hyperref}

\usepackage{epstopdf}

\definecolor{cream}{RGB}{222,217,201}

\begin{document}

\pagestyle{fancy}
\thispagestyle{plain}
\fancypagestyle{plain}{
\renewcommand{\headrulewidth}{0pt}
}

\makeFNbottom
\makeatletter
\renewcommand\LARGE{\@setfontsize\LARGE{15pt}{17}}
\renewcommand\Large{\@setfontsize\Large{12pt}{14}}
\renewcommand\large{\@setfontsize\large{10pt}{12}}
\renewcommand\footnotesize{\@setfontsize\footnotesize{7pt}{10}}
\renewcommand\scriptsize{\@setfontsize\scriptsize{7pt}{7}}
\makeatother

\renewcommand{\thefootnote}{\fnsymbol{footnote}}
\renewcommand\footnoterule{\vspace*{1pt}%
\color{cream}\hrule width 3.5in height 0.4pt \color{black} \vspace*{5pt}} 
\setcounter{secnumdepth}{5}

\makeatletter 
\renewcommand\@biblabel[1]{#1}            
\renewcommand\@makefntext[1]%
{\noindent\makebox[0pt][r]{\@thefnmark\,}#1}
\makeatother 
\renewcommand{\figurename}{\small{Fig.}~}
\sectionfont{\sffamily\Large}
\subsectionfont{\normalsize}
\subsubsectionfont{\bf}
\setstretch{1.125} 
\setlength{\skip\footins}{0.8cm}
\setlength{\footnotesep}{0.25cm}
\setlength{\jot}{10pt}
\titlespacing*{\section}{0pt}{4pt}{4pt}
\titlespacing*{\subsection}{0pt}{15pt}{1pt}

\fancyfoot{}
\fancyfoot[LO,RE]{\vspace{-7.1pt}\includegraphics[height=9pt]{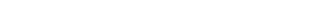}}
\fancyfoot[CO]{\vspace{-7.1pt}\hspace{13.2cm}\includegraphics{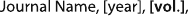}}
\fancyfoot[CE]{\vspace{-7.2pt}\hspace{-14.2cm}\includegraphics{RF}}
\fancyfoot[RO]{\footnotesize{\sffamily{1--\pageref{LastPage} ~\textbar  \hspace{2pt}\thepage}}}
\fancyfoot[LE]{\footnotesize{\sffamily{\thepage~\textbar\hspace{3.45cm} 1--\pageref{LastPage}}}}
\fancyhead{}
\renewcommand{\headrulewidth}{0pt} 
\renewcommand{\footrulewidth}{0pt}
\setlength{\arrayrulewidth}{1pt}
\setlength{\columnsep}{6.5mm}
\setlength\bibsep{1pt}

\makeatletter 
\newlength{\figrulesep} 
\setlength{\figrulesep}{0.5\textfloatsep} 

\newcommand{\topfigrule}{\vspace*{-1pt}%
\noindent{\color{cream}\rule[-\figrulesep]{\columnwidth}{1.5pt}} }

\newcommand{\botfigrule}{\vspace*{-2pt}%
\noindent{\color{cream}\rule[\figrulesep]{\columnwidth}{1.5pt}} }

\newcommand{\dblfigrule}{\vspace*{-1pt}%
\noindent{\color{cream}\rule[-\figrulesep]{\textwidth}{1.5pt}} }

\makeatother

\twocolumn[
  \begin{@twocolumnfalse}
{\includegraphics[height=30pt]{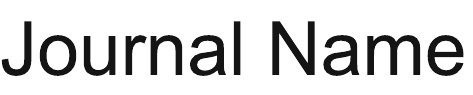}\hfill\raisebox{0pt}[0pt][0pt]{\includegraphics[height=55pt]{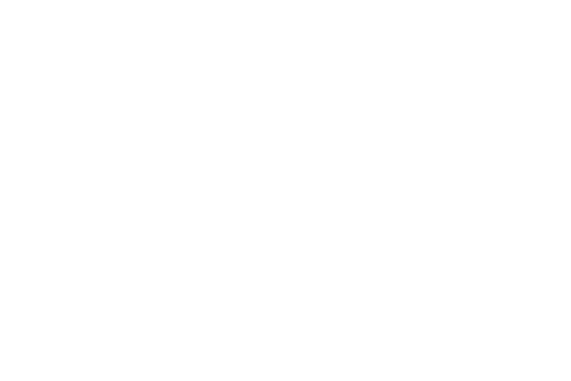}}\\[1ex]
\includegraphics[width=18.5cm]{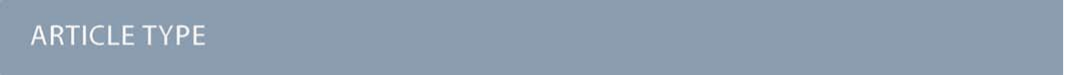}}\par
\vspace{1em}
\sffamily
\begin{tabular}{m{4.5cm} p{13.5cm} }

\includegraphics{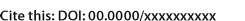} & \noindent\LARGE{\textbf{Analytical Modelling of the Transport in Analog Filamentary Conductive-Metal-Oxide/HfO$_{\rm x}$ ReRAM Devices$\dag$}} \\
 & \vspace{0.3cm} \\

 & \noindent\large{Donato Francesco Falcone,\textit{$^{\rm a}$} Stephan Menzel,\textit{$^{\rm b}$} Tommaso Stecconi,\textit{$^{\rm a}$} Matteo Galetta,\textit{$^{\rm a}$} Antonio La Porta,\textit{$^{\rm a}$} Bert Jan Offrein\textit{$^{\rm a}$} and Valeria Bragaglia\textit{$^{\rm a}$}} \\



\includegraphics{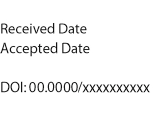} & \\

\end{tabular}

 \end{@twocolumnfalse} \vspace{0.6cm}

  ]

\renewcommand*\rmdefault{bch}\normalfont\upshape
\rmfamily
\section*{}
\vspace{-1cm}

\footnotetext{\textit{$^{\rm a}$~IBM Research Europe - Zürich, 8803 Rüschlikon, Switzerland. E-mail: dof@zurich.ibm.com}}
\footnotetext{\textit{$^{\rm b}$~Peter Gruenberg Institute 7, Forschungszentrum Juelich GmbH, 52425 Juelich, Germany. }}
\footnotetext{\dag~Electronic Supplementary Information (ESI) available. See DOI: 00.0000/00000000.}




\sffamily{\textbf{The recent co-optimization of memristive technologies and programming algorithms enabled neural networks training with in-memory computing systems. In this context, novel analog filamentary conductive-metal-oxide (CMO)/HfO$_{\rm x}$ redox-based resistive switching memory (ReRAM) represents a key technology. Despite device performance enhancements reported in literature, the underlying mechanism behind resistive switching is not fully understood. This work presents the first physics-based analytical model of the current transport and of the resistive switching in these devices. As a case study, analog TaO$_{\rm x}$/HfO$_{\rm x}$ ReRAM devices are considered. The current transport is explained by a trap-to-trap tunneling process, and the resistive switching by a modulation of the defect density within the sub-band of the TaO$_{\rm x}$ that behaves as electric field and temperature confinement layer. The local temperature and electric field distributions are derived from the solution of the electric and heat transport equations in a 3D finite element ReRAM model. The intermediate resistive states are described as a gradual modulation of the TaO$_{\rm x}$ defect density, which results in a variation of its electrical conductivity. The drift-dynamics of ions during the resistive switching is analytically described, allowing the estimation of defect migration energies in the TaO$_{\rm x}$ layer. Moreover, the role of the electro-thermal properties of the CMO layer is unveiled. The proposed analytical model accurately describes the experimental switching characteristic of analog TaO$_{\rm x}$/HfO$_{\rm x}$ ReRAM devices, increasing the physical understanding and providing the equations necessary for circuit simulations incorporating this technology.}} \\



\rmfamily 


\section*{Introduction}
\begin{figure}[t]
\centering
  \includegraphics[height=5.1cm]{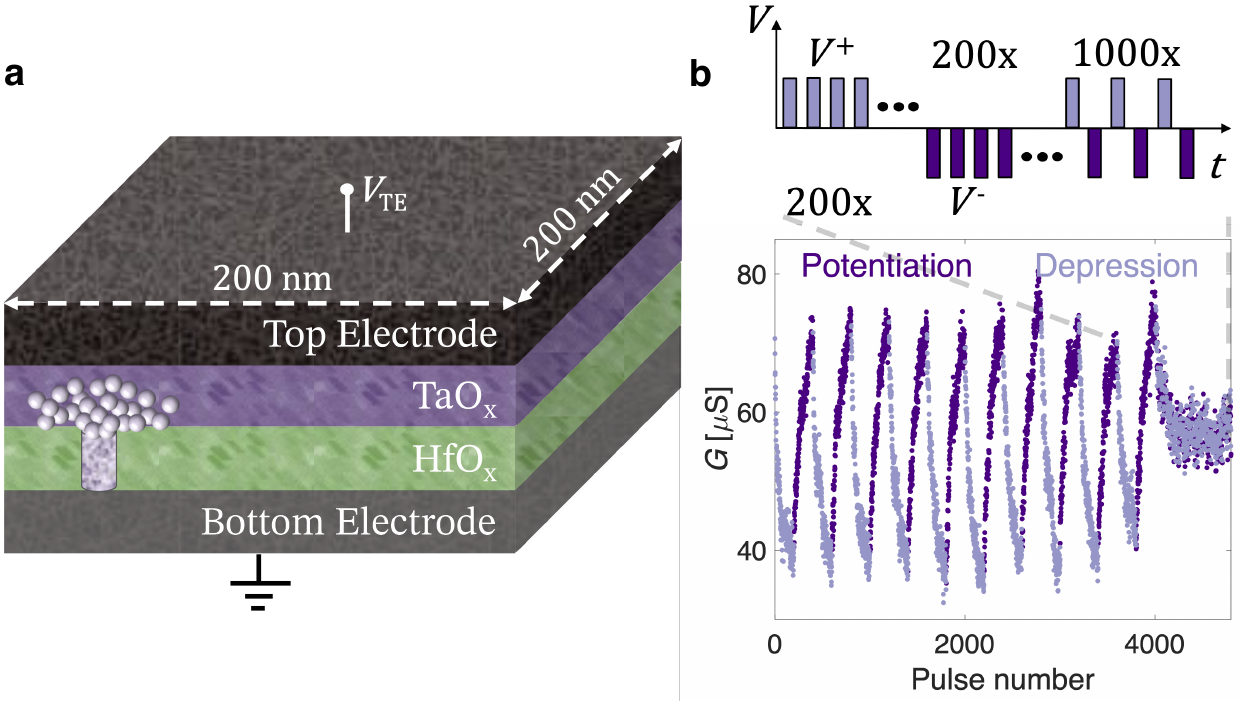}
  \caption{\textbf{(a)} Schematic illustration of the ReRAM device cross-section profile. \textbf{(b)} Bidirectional accumulative analog ReRAM response and symmetry point. The applied pulses have identical amplitudes of +1.6~V for the downward direction (\textit{G} depression) and -1.4~V for the upward direction (\textit{G} potentiation), each with a constant duration of 600~ns. After each pulse, the device’s resistance is measured by 10~$\rm \mu$s read pulses of $\pm$0.2~V. The top inset shows the electrical testing scheme utilized to evaluate the analog switching properties of the ReRAM device for the Tiki-Taka algorithm.}
  \label{Fig1}
\end{figure}

\textbf{Analog AI inference and training.} The recurrent data transfer between memory and processing units is a key source of speed and energy inefficiency in the hardware implementation of artificial neural networks (ANNs).\cite{Murmann2021} To address this challenge, modern digital AI accelerators enhance computational parallelism and bring memory closer to the processing units.\cite{Sebastian2020} However, due to the limited on-chip memory capacity and the increasing number of parameters in modern ANNs, communication with off-chip memory is still needed.\cite{Horowitz2014_4NanoTec,Jia2022_24_64coreAbu} To overcome this memory wall, a promising approach, known as analog in-memory computing (AIMC),\cite{Sebastian2019} consists of transitioning to a hybrid architecture where some of the arithmetic and logic operations can be performed at the location where the data is stored.\cite{Sebastian2017} Emerging resistive memory (memristive) technologies such as redox-based resistive switching memory (ReRAM),\cite{Yin2020_ReRAM} phase-change memory (PCM) \cite{PCM_HERMES} and magnetoresistive random access memory (MRAM) \cite{Deaville2022_MRAM}, are particularly well suited for the AIMC computing paradigm.\cite{Abu_64core} By organizing memristive devices in a crossbar configuration, matrix–vector multiplications (MVMs) can be performed directly in-memory by exploiting the multi-level storage capability of the devices, combined with Ohm’s law and Kirchhoff’s law, in O(1) time complexity.\cite{8_NanoTEC,Burr2017_AbuVLSI} AIMC-based AI hardware accelerator exploration has been mainly confined to inference workloads,\cite{Jain2023} being the accuracy of conventional training algorithms, such as the stochastic gradient descent (SGD), strongly dependent on the switching characteristics of the memristive devices.\cite{Gokmen2020} In particular, the key device requirement is the linear and symmetric conductance update in response to positive or negative pulse stimuli. However, since memristive devices exhibit non-ideal characteristics,\cite{Burr2016,Ielmini2016} advancements in both technology and training algorithms are required to achieve a full AIMC training system for deep learning workloads. Recent studies have demonstrated that parallel updates of AIMC ReRAM-based architecture may achieve a significant improvement in both energy efficiency and speed during neural network training with respect to digital AI accelerators.\cite{Gokmen2016,Gong2022} To address this challenge, a new ANN training algorithm, known as \textit{Tiki-Taka},\cite{Gokmen2020} was recently developed and optimized for analog resistive device. The core advantage of \textit{Tiki-Taka} is to limit the device symmetry requirements from the entire conductance (\textit{G}) window to a local symmetry point, where the changes in \textit{G} are equal in both upward and downward directions.\cite{Gokmen2020} The key device requirement for the \textit{Tiki-Taka} algorithm is analog resistive switching in both directions using a stream of identical pulses. To achieve so, the introduction of a properly engineered conductive-metal-oxide (CMO) in the conventional HfO$_{\rm x}$-based ReRAM Metal/Insulator/Metal (M/I/M) stack has been shown to lead to crucial device optimization.\cite{Kim2021} Such M/CMO/I/M ReRAM devices exhibit improved characteristics in terms of analog states, stochasticity, symmetry point and endurance, compared to conventional M/I/M technology,\cite{Gong2022} paving the way for AIMC training units using ReRAM crossbar arrays. \\
\textbf{Origin of resistive switching.} The origin of the resistive switching process in filamentary M/I/M HfO$_{\rm x}$-based ReRAM device is commonly attributed to the migration of oxygen vacancies in an oxygen-deficient conductive filament (CF).\cite{Ielmini2012} Such migration results in a purely filamentary conduction in the low resistance state (LRS) and tunneling transport through a low oxygen vacancy concentration region, known as gap, in the high resistance state (HRS).\cite{Ielmini2012} In contrast, the underlying mechanism behind the resistive switching in analog M/CMO/I/M HfO$_{\rm x}$-based ReRAM device is not fully understood, despite its device performance improvements. A recent study \cite{Falcone2023IMW} focuses on the electro-thermal modelling of static current transport in both LRS and HRS within a similar M/CMO/I/M ReRAM structure. However, the analytical current transport description and the switching dynamics, both crucial for a compact model description and consequently circuit simulations, remain unknown.\cite{Wu2017_TaOxTEL,Rambus_2014} \\
\textbf{Analog CMO/HfO$_{\rm x}$ ReRAM devices.} This work addresses this challenge by providing a physics-based analytical model of the conduction in analog M/CMO/I/M HfO$_{\rm x}$-based ReRAM devices. The electron transport and the drift ion dynamics are both analytically described in LRS, HRS, intermediate analog states and during the \textit{SET} (from HRS to LRS) and \textit{RESET} (from LRS to HRS) transitions. The model results are compared to experimental data measured on a representative TiN/TaO$_{\rm x}$/HfO$_{\rm x}$/TiN ReRAM device, with a cell area of (200~nm)$^{\rm 2}$ and the same stack as reported in our previous work.\cite{TEC2} Figure \ref{Fig1}a shows a simplified cross-sectional schematic of the analog TaO$_{\rm x}$/HfO$_{\rm x}$ ReRAM devices, in shared bottom electrode configuration, modelled in this work. The bidirectional accumulative ReRAM response under the application of trains of identical voltage pulses with alternating polarity in batches of 200, is illustrated in Figure \ref{Fig1}b. This pulse scheme is repeated 10 times, showing a stable \textit{G} window. Finally, by alternately applying a single up pulse and down pulse, a stable \textit{G} symmetry point is achieved. This process replicates the programming conditions utilized in \textit{Tiki-Taka} training.\cite{Gokmen2020} As reported in our previous works,\cite{TEC_ESSDERC,Stecconi2022} analog TaO$_{\rm x}$/HfO$_{\rm x}$ ReRAM devices are characterized by $I_{\rm SET} \approx I_{\rm RESET}$ and consequently $V_{\rm SET} \approx V_{\rm RESET}$. However, to increase the switching ratio ($R_{\rm HRS}$/$R_{\rm LRS}$), a larger stop voltage amplitude is desirable for the \textit{RESET} compared to the \textit{SET} process ($|V_{\rm stop,RESET}|$ > $|V_{\rm stop,SET}|$). The fabrication process is detailed in our previous work,\cite{TEC2} while additional information on the electrical characterization of ReRAM is reported in the \textit{Methods} section of the Supplementary Information, ESI \dag.
Although in \textit{Tiki-Taka} training applications the targeted programming scheme employs short pulses (from hundreds of ps \cite{Abedin2023} to $\rm \mu$s \cite{TEC2}), quasi-static current-voltage (\textit{I}-\textit{V}) characterization of the ReRAM device is required to model the physics of the steady-state current transport. However, by cycling the TaO$_{\rm x}$/HfO$_{\rm x}$ devices with quasi-static voltage excitations, intentionally without current compliance control to prevent masking device physics during the switching, the conductance window shifts toward larger \textit{G} values by a factor of around 3, as shown in Figures \ref{Fig23}b and \ref{Fig23}d. Nevertheless, the physics-based analytical current transport model developed in this work and applied to such TaO$_{\rm x}$/HfO$_{\rm x}$ devices, can also be used to explain other similar analog M/CMO/I/M ReRAM systems, including those with lower average conductance.
\section*{Physics-based analytical model}
\textbf{Current transport: the model.} In sub-stoichiometric metal-oxides, a significant concentration of defects related to oxygen vacancies and metal interstitials is observed.\cite{Bao2023,Perevalov2007_32Kefei,Zhu2016} Each defect forms isolated donor-like states in the band gap of the metal-oxide, which become a sub-band for large defect concentrations.\cite{Bao2023,Funck2021, Goldfarb2014,Graves2017} In oxides with a high band gap, there can be a substantial energetic separation (activation energy) between the defect states and the lower edge of the conduction band $E_{\rm CB}$. For example, in TaO$_{\rm x}$, the spectral location of the defect state sub-band associated with oxygen vacancies has been reported to be in the midgap, approximately 2~eV above the top of the valence band $E_{\rm VB}$.\cite{Goldfarb2014} Consequently, according to the energy band alignment between the Fermi level of the electrodes and the defect sub-band in the conductive metal-oxide, it is likely that electrons coming from one electrode are trapped in the defect states during the current transport. However, to actively contribute to the conduction, trapped electrons must be released. Depending on the defect activation energy in the conductive metal-oxide, two scenarios may occur: the electrons get released by thermal excitation in the conduction band, or they tunnel from trap-to-trap until reaching the other electrode. In this work, due to midgap electronic states in the TaO$_{\rm x}$ layer, the electronic current $I_{\rm e}$ in a filamentary HfO$_{\rm x}$/TaO$_{\rm x}$ ReRAM device is modelled as a trap-to-trap tunneling process in the TaO$_{\rm x}$ layer, based on the description provided by Mott and Gurney. \cite{Mott_Gurney1950} It describes electron-hopping conduction over an energy barrier $\Delta E_{\rm e}$, which is identical in every direction in the absence of an electric field. Application of an electric field modifies the energy barrier by $\mp$ $ea_{\rm e}E$/2 for forward (backward) jumps, resulting in a lowered (raised) barrier. 
\begin{equation}
   I_{\rm e}^{\rm Mott-Gurney} = 2 A e a_{\rm e} \nu_{\rm 0,e} N_{\rm e} \exp{(\frac{-\Delta E_{\rm e}}{k_{\rm B} T})} \sinh{(\frac{a_{\rm e}eE}{2k_{\rm B} T})}
   \label{I_e_MG}
\end{equation}
In equation \eqref{I_e_MG}, $e$ is the elementary charge, $k_{\rm B}$ is the Boltzmann’s constant, $a_{\rm e}$ is the hopping distance, $\nu_{\rm 0,e}$ is the electron attempt frequency, $N_{\rm e}$ is the density of electronic defect states in the TaO$_{\rm x}$ band gap, $\Delta E_{\rm e}$ is the zero-field hopping energy barrier, $T$ and $E$ are the local temperature and electric field, respectively, and $A = \rm \pi \it r_{\rm CF}^2$, $r_{\rm CF}$ being the filament radius, is the cross-sectional area of the filament at the interface with the TaO$_{\rm x}$ layer. Among the different types of defects in the TaO$_{\rm x}$ layer, this work focuses on oxygen vacancy defects $V_{\rm O}^{\rm \cdot \cdot}$, which have been reported to generally have lower formation and migration energies with respect to Ta$_{\rm i}$ interstitials.\cite{Perevalov2007_32Kefei,Zhu2016} Each double-positively charged vacancy $V_{\rm O}^{\rm \cdot \cdot}$ generates two electronic defect states in the band gap.\cite{Zhu2016} Therefore, the maximum density of defect states $N_{\rm e}^{\rm max}$ that may contribute to the transport can be related to the ionic vacancy concentration $N_{\rm V_{\rm O}^{\rm \cdot \cdot}}$, according to $ N_{\rm e}^{\rm max} = 2 N_{\rm V_{\rm O}^{\rm \cdot \cdot}}$. Depending on the energy alignment between the defect sub-band in the TaO$_{\rm x}$ layer and the Fermi level of the electrodes, statistically only a sub-set of the defect states $N_{\rm e}$ contributes to the electronic transport, resulting in equation \eqref{Nemax_vs_Nion}.
\begin{equation}
   N_{\rm e} = \beta N_{\rm e}^{\rm max} = 2 \beta N_{\rm V_{\rm O}^{\rm \cdot \cdot}}
   \label{Nemax_vs_Nion}
\end{equation}
where $\beta$ is a scaling factor between 0 and 1 to filter out the defect states that are not contributing to the conduction. \\
\textbf{Defect drift dynamics: the model.} In addition to electronic transport, the description of the ion dynamics is crucial for understanding the resistive switching process and the resulting structural material changes. To do so, the drift dynamics of the ionic defects $V_{\rm O}^{\rm \cdot \cdot}$ in the TaO$_{\rm x}$ layer during the \textit{SET} and \textit{RESET} processes is similarly described by an ion-hopping conduction over a migration energy barrier $\Delta E_{\rm ion}$,\cite{Camilla2019} according to equations \eqref{Iion_drift1} and \eqref{Iion_drift2},
\begin{equation}
   v_{\rm ion, drift}^{\rm Mott-Gurney} = 2 a_{\rm ion} \nu_{\rm 0,ion} \exp{(\frac{-\Delta E_{\rm ion}}{k_{\rm B} T})} \sinh{(\frac{a_{\rm ion}z_{\rm V_{O}^{\cdot \cdot}}eE}{2k_{\rm B} T})}
   \label{Iion_drift1}
\end{equation}
\begin{equation}
   I_{\rm ion} = A e z_{\rm V_{O}^{\cdot \cdot}} N_{\rm V_{\rm O}^{\rm \cdot \cdot}} v_{\rm ion, drift}^{\rm Mott-Gurney}
   \label{Iion_drift2}
\end{equation}
where $v_{\rm ion, drift}^{\rm Mott-Gurney}$ is the ion drift velocity, $z_{\rm V_{O}^{\cdot \cdot}}$ is the charge number of the oxygen vacancies relative to the perfect crystal, $a_{\rm ion}$ is the crystal hopping distance and $\nu_{\rm 0,ion}$ is the ion attempt frequency. \\
\textbf{3D Finite Element Model (FEM).} The local temperature $T$ and electric field $E$ distributions in the steady state of TaO$_{\rm x}$/HfO$_{\rm x}$ ReRAM devices are assessed using a space resolved 3D FEM developed with COMSOL Multiphysics 5.2 software, according to equations \eqref{Comsol_eq1} and \eqref{Comsol_eq2}.
\begin{equation}
  \nabla \cdot (\sigma (-\nabla V) = 0
  \label{Comsol_eq1}
\end{equation}
\begin{equation}
   \nabla \cdot (- k \nabla T) = J_{\rm e} \cdot E = Q_{\rm e}
   \label{Comsol_eq2}
\end{equation}
where $J_{\rm e}$ is the electric current density, $\sigma$ the electrical conductivity, $V$ the electric potential, $k$ the thermal conductivity and $Q_{\rm e}$ the heat source due to Joule heating. Table \ref{Table1} lists the basic simulations parameters used in this work, taken from literature.
\begin{table}[h]
\small
  \caption{\ Simulation parameters from literature}
  \label{tbl:example}
  \begin{tabular*}{0.48\textwidth}{@{\extracolsep{\fill}}llll}
    \hline
    Symbol & Value & Symbol & Value \\
    \hline
    $\nu_{\rm 0,ion}$ & 4 $\cdot$ 10$^{12}$~Hz \cite{Camilla2019} & $\nu_{\rm 0,e}$ & 2 $\cdot$ 10$^{13}$~Hz \cite{Lew2020} \\
    $a_{\rm ion}$ & 0.4~nm \cite{Hur2019} & $k_{\rm TaO_x}$ & 1~W(mK)$^{-1}$ \cite{kTaOx} \\
    $k_{\rm HfO_x}$ & 0.5~W(mK)$^{-1}$ \cite{k_HfOx} & $k_{\rm CF}$ & 23~W(mK)$^{-1}$ \cite{ThermalConductivityHf} \\
    $k_{\rm TiN,Sputter}$ & 3~W(mK)$^{-1}$ \cite{k_TiN_Sputter} & $k_{\rm TiN,ALD}$ & 11~W(mK)$^{-1}$ \cite{k_TiN_ALD}\\
    $\sigma_{\rm HfO_x}$ & 1 $\cdot$~10$^{-4}$ Sm$^{-1}$ & $\sigma_{\rm CF}$ & 4.2 $\cdot$~10$^{4}$ Sm$^{-1}$ \cite{Falcone2023IMW} \\
    $\sigma_{\rm TiN}$ & 5 $\cdot$~10$^{5}$ Sm$^{-1}$ \cite{TiN_sigma} &sweep rate & 0.1~V/s\\
    $T_{\rm 0}$ & 293~K &  $z_{\rm V_{O}^{\cdot \cdot}}$ & 2\\
    \hline
  \end{tabular*}
  \label{Table1}
\end{table}\\
The boundary conditions include the current conservation to all the domains, and electric isolation ($\rm n$ $\cdot$ $J_{\rm e} = 0$), where $\rm n$ is the outward normal vector, to all the vertical external boundaries such that no electric current flows through the boundaries. The electric potential is always applied to the ReRAM device top electrode, with the bottom electrode serving as the ground node, as shown in Figure \ref{Fig1}a. In the 3D FEM, all external boundaries of the ReRAM device are maintained at room temperature $T_{\rm 0}$ in steady state, as negligible heating is expected far from the switching area. The model takes into account the entire experimental TaO$_{\rm x}$/HfO$_{\rm x}$ ReRAM device area of (200~nm)$^{\rm 2}$.




\section*{Results and discussion}
\textbf{Analytical model of the LRS and HRS.} An high-resolution scanning transmission electron microscopy (STEM) measurement on a similar TaO$_{\rm x}$/HfO$_{\rm x}$ ReRAM stack unveils a thin oxygen-rich TaO$_{\rm x}$ amorphous phase ($\approx$ 3~nm) between the crystalline oxygen-deficient TaO$_{\rm x}$ and the HfO$_{\rm x}$ layer.\cite{Stecconi2022,TEC_ESSDERC} The amorphous TaO$_{\rm x}$ phase, combined with the reduced oxygen scavenging capabilities of TaO$_{\rm x}$ material with respect to Ti,\cite{KimMenzel2016,Goux2014VLSI} results in an increase of the forming voltage $V_{\rm forming}$ in the TaO$_{\rm x}$/HfO$_{\rm x}$ ReRAM device ($\approx$ 3.6~V)\cite{TEC2} compared to Ti/HfO$_{\rm x}$ technology ($\approx$ 2.5~V)\cite{Stecconi2022}. During the forming, a dielectric breakdown occurs in the HfO$_{\rm x}$ and oxygen-rich amorphous TaO$_{\rm x}$ layers, leading to the formation of a highly defect-rich conductive filament in both layers.\cite{TEC_ESSDERC} For oxygen vacancy defects, reported formation energies range from 2.8~eV to 4.6~eV in HfO$_{\rm x}$, depending on the stoichiometry,\cite{Padovani2012,Padovani2013} and from 4.3~eV to 5.7~eV in TaO$_{\rm x}$, depending on the oxygen vacancy charge number.\cite{32ZhuLee2013,40ZhuGuo2014} 
\begin{figure*}[t]
 \centering
 \includegraphics[height=11cm]{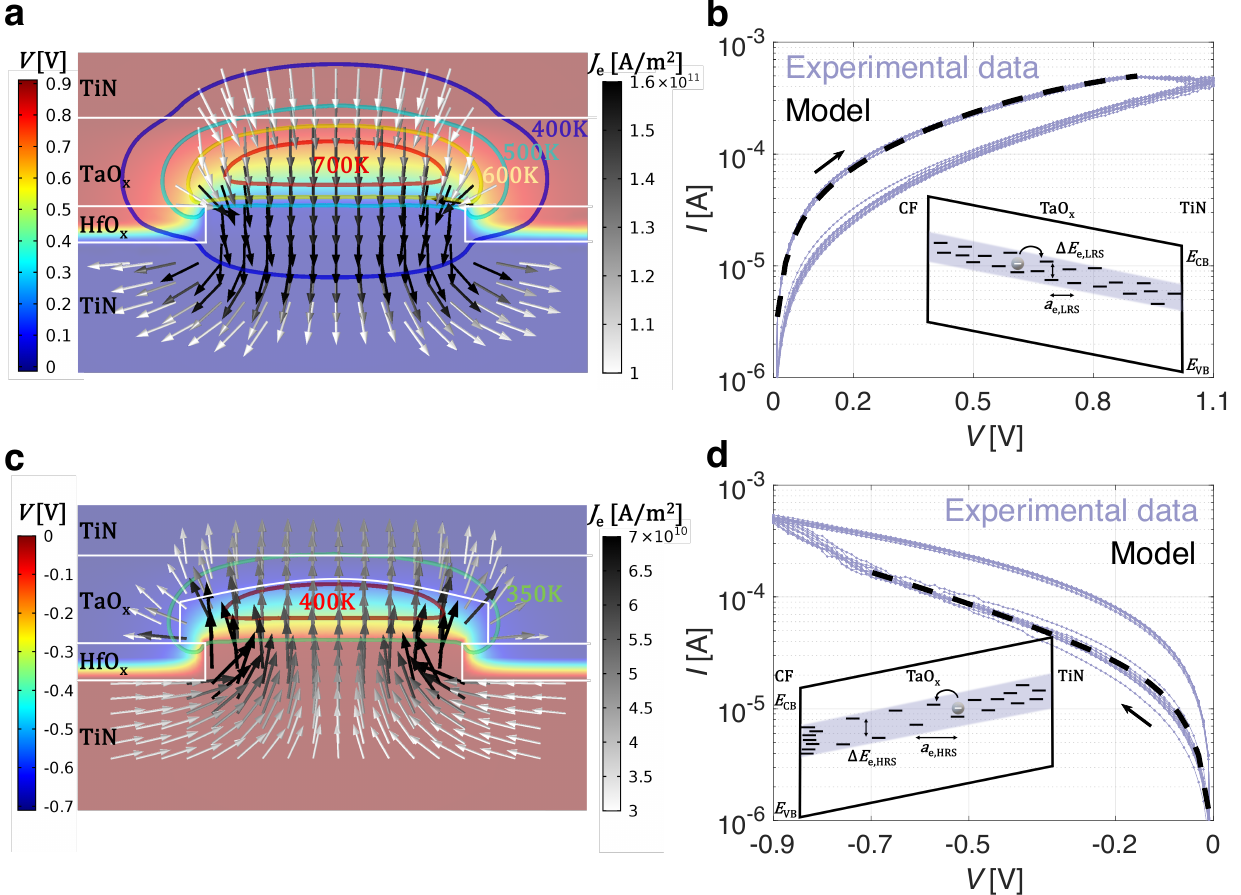}
 \caption{The steady-state 2D spatial distributions of \textit{T}, \textit{V} and \textit{J$_{\rm e}$} in a TaO$_{\rm x}$/HfO$_{\rm x}$ ReRAM device is shown \textbf{(a)} in LRS with a constant bias of 0.9~V and \textbf{(c)} in HRS with a constant bias of -0.7~V. Both \textit{T} and \textit{V} are locally confined in a 3D half-spherical volume in the TaO$_{\rm x}$ layer. The key purpose of the CF is the \textit{J$_{\rm e}$} confinement. The electro-thermal energy confinement in the TaO$_{\rm x}$ layer leads to the \textbf{(a)} \textit{RESET} or \textbf{(c)} \textit{SET} switching process. The experimental quasi-static \textit{I}-\textit{V} characteristic of a TaO$_{\rm x}$/HfO$_{\rm x}$ ReRAM device and the analytical electron trap-to-trap tunneling model are reported \textbf{(b)} in LRS and \textbf{(d)} HRS. The bottom insets in \textbf{(b)} and \textbf{(d)} show a schematic sketch of the trap-to-trap tunneling transport in the TaO$_{\rm x}$ sub-band, considering the CF and the TiN as electrodes, in LRS and HRS respectively.}
 \label{Fig23}
\end{figure*}
Due to the high formation energy, defect generation, both in HfO$_{\rm x}$ and TaO$_{\rm x}$ materials, occurs with statistical relevance only during the forming, when a large electric field and high temperature are locally reached. Therefore, the resistive switching process in a TaO$_{\rm x}$/HfO$_{\rm x}$ ReRAM device is determined by the migration of pre-existing defects, which leads to the modulation of the density of defect states in the band gap of the TaO$_{\rm x}$ layer. A 3D FEM is developed to simulate the electronic transport in a filamentary TaO$_{\rm x}$/HfO$_{\rm x}$ ReRAM device in both LRS and HRS. The electrical conductivity of the crystalline sputtered TaO$_{\rm x}$ layer is measured experimentally.\cite{Stecconi2022} However, due to the defect generation during the forming event, the local conductivity of the TaO$_{\rm x}$ layer on top of the conductive filament will be higher. For this reason, the electrical conductivity of the crystalline TaO$_{\rm x}$ layer was considered as a parameter in the 3D ReRAM model, within experimentally reported boundaries.\cite{Stecconi2022} From the model of the electronic conduction of the TaO$_{\rm x}$/HfO$_{\rm x}$ ReRAM device in the low-voltage linear regime in LRS, both the filament radius $r_{\rm CF}$ and the local conductivity of the TaO$_{\rm x}$ layer $\sigma_{\rm TaO_{\rm x}}$ are extracted and reported in Table \ref{Table2}. Figure \ref{Fig23}a shows the steady-state 2D spatial distributions of temperature, electric potential, and current density in a TaO$_{\rm x}$/HfO$_{\rm x}$ ReRAM device in LRS with a constant bias of 0.9~V, taking into account the experimental \textit{I}-\textit{V} non-linearity with the same approach as reported in our previous work.\cite{Falcone2023IMW} The TaO$_{\rm x}$ layer, with its lower thermal conductivity (see Table \ref{Table1}), represents a larger thermal resistance compared to the conductive filament. For this reason, in steady state, a spatial temperature gradient exists within the TaO$_{\rm x}$ layer (see Figure \ref{Fig23}a), acting as a thermal confinement layer,\cite{Wu2017_TaOxTEL} confining heat on top of the conductive filament. In addition, due to the combination of the filamentary current constriction and the electrical conductivity of the TaO$_{\rm x}$ layer, the resistance in the LRS is dominated by the spreading resistance ($R_{\rm spreading} \propto (\rm 2\pi \it \sigma_{\rm TaO_x} \it r_{\rm CF} )^{-1} \gg$ $R_{\rm CF}$) between the CF and the TaO$_{\rm x}$, which leads to a nonuniform current density in the TaO$_{\rm x}$ layer.\cite{Falcone2023IMW} As a consequence, by properly designing its electrical conductivity,\cite{Falcone2023IMW} the TaO$_{\rm x}$ layer can also act as a field confinement layer, confining the applied electric field above the conductive filament in a half-spherical shape, as illustrated in Figure \ref{Fig23}a. Although the thermal and electric field confinement can similarly be achieved with an insulating metal-oxide layer, to avoid an increase of the overall forming voltage, a conductive TaO$_{\rm x}$ is used. The average temperature and electric field over such a 3D half-spherical volume in the TaO$_{\rm x}$ layer are calculated (see Supplementary Information Figure S1, ESI \dag), and used in equation \ref{I_e_MG} to describe the static electronic conduction in LRS. Figure \ref{Fig23}b shows the experimental clockwise \textit{I}-\textit{V} quasi-static characteristic of a representative TaO$_{\rm x}$/HfO$_{\rm x}$ ReRAM device in LRS over 10 cycles. The analytical electron trap-to-trap tunneling model in the TaO$_{\rm x}$ layer (see Figure \ref{Fig23}b) accurately describes the experimental curve in LRS, enabling the extraction of $a_{\rm e,LRS}$, $\Delta E_{\rm e,LRS}$, and $N_{\rm e,LRS}$, as reported in Table \ref{Table2}. Oxygen vacancy migration in the TaO$_{\rm x}$ layer has been reported as a process with an activation energy of $\approx$ 1.5~eV.\cite{Zhu2016,Hur2019} As the applied positive electric potential on the TaO$_{\rm x}$/HfO$_{\rm x}$ ReRAM device increases, the electro-thermal energy in the TaO$_{\rm x}$ layer rises, through the local confinement of both the electric field and temperature. When the oxygen vacancy migration energy in the TaO$_{\rm x}$ layer is reached, the \textit{RESET} switching process starts. 
During the \textit{RESET} process, oxygen vacancies in the TaO$_{\rm x}$ layer radially migrate toward the conductive filament. Depending on the energy barrier at the TaO$_{\rm x}$/HfO$_{\rm x}$ interface, two scenarios may occur: an exchange of oxygen ions between the two layers, or an accumulation of oxygen vacancies within the TaO$_{\rm x}$ layer at the interface with the conductive filament.\cite{Sommer2023} Both scenarios lead to a high resistance state (HRS) in the TaO$_{\rm x}$/HfO$_{\rm x}$ ReRAM device, characterized by a partial defect depletion of a half-spherical volume, also referred to as a dome \cite{Falcone2023IMW}, in the TaO$_{\rm x}$ layer. In this work, although it's not a binding assumption for the validity of the model, the resistive change of the TaO$_{\rm x}$/HfO$_{\rm x}$ ReRAM device is primarily attributed to oxygen vacancy migration within the TaO$_{\rm x}$ layer. In agreement with this interpretation, a 3D TaO$_{\rm x,Dome}$ volume ($V_{\rm TaO_x,Dome}$), characterized by a lower density of defect states and therefore lower electrical conductivity $\sigma_{\rm TaO_{\rm x,Dome}}$ (see Table \ref{Table2}), has been employed to model electronic conduction in TaO$_{\rm x}$/HfO$_{\rm x}$ ReRAM devices in HRS. Figure \ref{Fig23}c shows the steady-state 2D spatial distributions of temperature, electric potential, and current density in a TaO$_{\rm x}$/HfO$_{\rm x}$ ReRAM device in HRS with a constant bias of -0.7~V, taking into account the experimental \textit{I}-\textit{V} non-linearity with the same approach as reported in our previous work.\cite{Falcone2023IMW} Similar to the LRS, TaO$_{\rm x}$ acts as both a thermal and field confinement layer (see Figure \ref{Fig23}c). The average temperature and electric field (see Supplementary Information Figure S1, ESI \dag) serve as inputs in equation \ref{I_e_MG} to model the static current transport in HRS. The same analytical trap-to-trap tunneling model accurately describes the experimental curve of a TaO$_{\rm x}$/HfO$_{\rm x}$ ReRAM device also in HRS (see Figure \ref{Fig23}d). The physical parameters characterizing the trap-to-trap tunneling process in the TaO$_{\rm x,Dome}$, $a_{\rm e,HRS}$, $\Delta E_{\rm e,HRS}$, and $N_{\rm e,HRS}$ are reported in Table \ref{Table2}. Compared to the LRS, the less-defective TaO$_{\rm x,Dome}$ volume is characterized by both a larger hopping distance $a_{\rm e}$ and energy barrier $\Delta E_{\rm e}$, and a lower density of defect states $N_{\rm e}$ in the band gap. By increasing the applied negative electric potential, when the required activation energy is reached, the \textit{SET} switching process starts. Oxygen vacancies accumulated in the TaO$_{\rm x}$ layer at the interface with the conductive filament, radially migrate back and repopulate the half-spherical TaO$_{\rm x,Dome}$, resulting in the LRS of the TaO$_{\rm x}$/HfO$_{\rm x}$ ReRAM device.
\begin{table}[h]
\small
  \caption{\ Simulation results from our model}
  \label{tbl:example}
  \begin{tabular*}{0.48\textwidth}{@{\extracolsep{\fill}}llll}
    \hline
    Symbol & Value & Symbol & Value \\
    \hline
    $r_{\rm CF}$ & 25~nm & $V_{\rm TaO_x,Dome}$ & 3 $\cdot$ 10$^{-23}$~m$^{3}$ \\
    $\sigma_{\rm TaO_x}$ & 2 $\cdot$ 10$^{3}$~Sm$^{-1}$ & $\sigma_{\rm TaO_x,Dome}$ & 0.38 $\cdot$ 10$^{3}$~Sm$^{-1}$ \\
    $a_{\rm e,LRS}$ & 0.75~nm & $a_{\rm e,HRS}$ & 1.1~nm \\
    $\Delta E_{\rm e,LRS}$ & 55~meV & $\Delta E_{\rm e,HRS}$ & 88~meV \\
    $N_{\rm e,LRS}$ & 3.6 $\cdot$ 10$^{26}$~m$^{-3}$ & $N_{\rm e,HRS}$ & 1.7 $\cdot$ 10$^{26}$~m$^{-3}$ \\
    \hline
  \end{tabular*}
  \label{Table2}
\end{table}\\
\textbf{Intermediate resistive states.} By applying bipolar short pulse streams, the conductance of TaO$_{\rm x}$/HfO$_{\rm x}$ ReRAM devices can be gradually potentiated or depressed, as shown in Figure \ref{Fig1}b. The number of states and the overall analog switching performance have been shown to improve upon the application of ultrafast pulses (down to 300~ps).\cite{Abedin2023} Additionally, in our previous work,\cite{TEC2} it was demonstrated that TaO$_{\rm x}$/HfO$_{\rm x}$ ReRAM devices preserve analog properties after more than 10$^{7}$ programming pulses, and show stable intermediate resistive states (<4\% drift after 72~h at 85~\textsuperscript{o}C). Due to the excellent endurance and retention properties of TaO$_{\rm x}$/HfO$_{\rm x}$ ReRAM devices, in first approximation, device non-idealities are not included in this work. The 3D FEM of the TaO$_{\rm x}$/HfO$_{\rm x}$ ReRAM device was used to model the intermediate resistive states resulting from the application of quasi-static excitations with increasing amplitudes. Figure \ref{Fig45}a shows the experimental response of a representative TaO$_{\rm x}$/HfO$_{\rm x}$ ReRAM device to quasi-static voltage sweeps with an increasing stop voltage amplitude $V_{\rm stop}$. Even with quasi-static excitations, intermediate resistive states can be accessed during both the \textit{SET} and the \textit{RESET}, without any external compliance current control. Figures \ref{Fig45}b and \ref{Fig45}c illustrate a multistate exponential relationship between the experimental resistance measured at 0.2~V and $V_{\rm stop}$ during the \textit{SET} and \textit{RESET} processes, respectively. A comparable trend during the \textit{RESET} of a filamentary ReRAM device has already been reported,\cite{Lew2020} but not for the \textit{SET} process, which is typically a self-accelerated abrupt process.\cite{Ielmini2012} In agreement with the interpretation provided in this work (see Figure \ref{Fig23}), the equivalent electrical conductivity $\sigma_{\rm TaO_{\rm x,Dome}}$, assumed of a constant volume $V_{\rm TaO_{\rm x,Dome}}$, to model the electronic conduction in the low-voltage linear regime (see Supplementary Information Figure S2, ESI \dag), is extracted. Figures \ref{Fig45}b and \ref{Fig45}c show the exponential evolution of the $\sigma_{\rm TaO_{\rm x,Dome}}$ as a function of $V_{\rm stop}$ during the \textit{SET} and \textit{RESET} processes, respectively. \\
\textbf{Analytical model of the switching characteristic.} To model the current during the entire switching characteristic, in addition to the LRS and HRS (see Figure \ref{Fig23}), the transport during the \textit{SET} and \textit{RESET} transitions has to be addressed. To do so, the analytical trends of the electro-thermal conditions and trap-to-trap tunneling parameters during such transitions, are required. In Figures \ref{Fig45}b and \ref{Fig45}c, both resistance, based on experimental evidence, and $\sigma_{\rm TaO_{\rm x,Dome}}$, based on model evidence, exhibit an exponential dependence on $V_{\rm stop}$ in the TaO$_{\rm x}$/HfO$_{\rm x}$ ReRAM devices. Additionally, the electro-thermal conditions (see Supplementary Information Figure S1, ESI \dag) and trap-to-trap tunneling parameters (see $a_{\rm e}$, $\Delta E_{\rm e}$ and $N_{\rm e}$ from Table \ref{Table2}) are known at the onset of both the \textit{SET} and \textit{RESET} processes from the modelling of the static electronic transport. By combining these evidences, an exponential distribution for $T$, $E$, $a_{\rm e}$, $\Delta E_{\rm e}$, and $N_{\rm e}$ within the static boundaries (from Figure \ref{Fig23} and Table \ref{Table2}) is assumed to model the electronic current during both the \textit{SET} and \textit{RESET} transitions. Figure \ref{Fig45}d shows the resulting electron trap-to-trap tunneling transport model during the \textit{SET} and \textit{RESET} transitions according to equation \ref{I_e_MG}. The model in Figure \ref{Fig45}d accurately describes the experimental quasi-static switching characteristic, validating the previous assumption. 
\begin{figure*}[t]
 \centering
 \includegraphics[height=9.3cm]{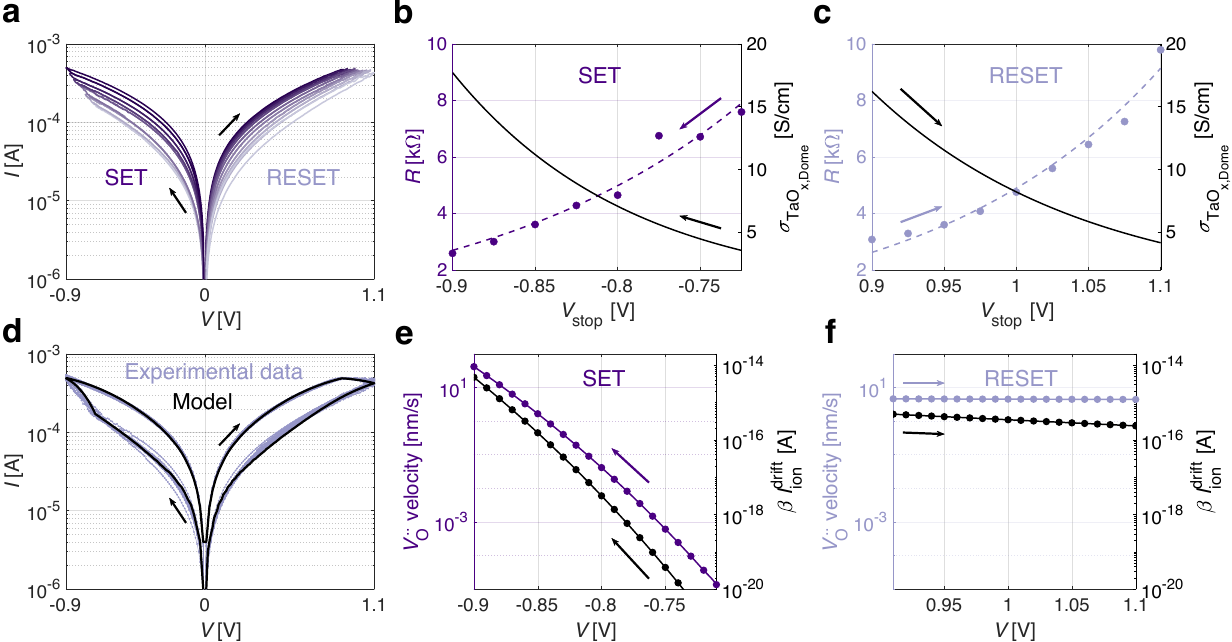}
 \caption{(a) Experimental quasi-static voltage sweeps, where the absolute stop voltage amplitude for both \textit{SET} (from -0.725~V to -0.9~V) and \textit{RESET} (from 0.9~V to 1.1~V) processes increases gradually with a step of 0.25~V. The exponential stop voltage-dependent evolution of the experimental intermediate resistive states measured at 0.2~V and the simulated equivalent electrical conductivity $\sigma_{\rm TaO_{\rm x,Dome}}$ are reported for \textit{SET} (b) and \textit{RESET} (c). (d) The analytical trap-to-trap tunneling model accurately describes the experimental switching characteristic of a TaO$_{\rm x}$/HfO$_{\rm x}$ ReRAM device. (e) Oxygen vacancy drift velocity and the resulting ionic current scaled by the $\beta$ factor, are reported as a function of the applied quasi-static voltage for \textit{SET} (e) and \textit{RESET} (f), respectively.}
 \label{Fig45}
\end{figure*}
During the \textit{RESET} transition, the temperature slightly decreases while the electric field in the TaO$_{\rm x}$ layer increases (see Supplementary Information Figure S3, ESI \dag). The simulation aligns with the interpretation presented in this work, indicating an overall increase in resistance, resulting in reduced Joule heating and a progressively increasing electric field confined within the less-defective TaO$_{\rm x,Dome}$. During the \textit{SET} transition, although the applied voltage increases, the electric field in the TaO$_{\rm x}$ layer stays nearly the same, while the temperature increases exponentially (see Supplementary Information Figure S3, ESI \dag). This aligns with a decrease in resistance, resulting in enhanced Joule heating and a defect-enrichment of the TaO$_{\rm x,Dome}$, restoring the LRS condition. \\
\textbf{$V_{\rm O}^{\rm \cdot \cdot}$ drift dynamics.} To explain the dynamics of the resistive switching in TaO$_{\rm x}$/HfO$_{\rm x}$ ReRAM devices, the drift of $V_{\rm O}^{\rm \cdot \cdot}$ during the \textit{SET} and \textit{RESET} transitions is modelled. For each voltage step (10~mV) during the \textit{SET} and \textit{RESET} processes, the average oxygen vacancy drift velocity is computed according to equation \ref{Iion_drift1}. By scaling the drift velocity with the experimental measurement time (100~ms/step), an average drift induced displacement is computed for each voltage step. The cumulative sum of the displacements during both the \textit{SET} and \textit{RESET} transitions, results in the average size of the dome along the vertical direction. For consistency with the current transport model (see Figure \ref{Fig23}), an overall oxygen vacancy displacement of approximately 10~nm is assumed for both the \textit{SET} and \textit{RESET}, while the ion-hopping energy barriers $\Delta E_{\rm ion,SET}$ and $\Delta E_{\rm ion,RESET}$ are considered as fitting parameters. The extracted ion activation energies are $\Delta E_{\rm ion,RESET}$ = 1.44~eV and $\Delta E_{\rm ion,SET}$ = 1.32~eV. Consistently, both energies closely align with the reported oxygen vacancy migration energy of $\approx$ 1.5~eV \cite{Zhu2016} in the TaO$_{\rm x}$ layer. The difference between $\Delta E_{\rm ion,RESET}$ and $\Delta E_{\rm ion,SET}$ may suggest that the two migrations happen through different energy paths. Oxygen vacancy accumulation at the interface with the conductive filament (during \textit{RESET}) may require more energy than redistribution back into the 3D half-spherical volume (during \textit{SET}). Figures \ref{Fig45}e and \ref{Fig45}f show the oxygen vacancy drift velocity and the resulting ionic current scaled by the $\beta$ factor, according to equation \ref{Iion_drift2}, during the quasi-static \textit{SET} and \textit{RESET}, respectively. Although the $\beta$ factor, which describes the sub-set of defect states contributing to conduction, may vary during the switching process, the change in $\beta I_{\rm ion}^{\rm drift}$ is primarily attributed to a change in $I_{\rm ion}^{\rm drift}$, which depends on the electro-thermal conditions according to equation \ref{Iion_drift2}. In agreement with the experimental switching characteristic, the ionic drift dynamics follow an exponentially increasing trend during \textit{SET} (see Figure \ref{Fig45}e) and a nearly uniform one during \textit{RESET} (see Figure \ref{Fig45}f). Future work will focus on establishing the analytical link between the $V_{\rm O}^{\rm \cdot \cdot}$ average drift velocity and the ionic vacancy concentration $N_{\rm V_{\rm O}^{\rm \cdot \cdot}}$, which determines the electronic current according to equations \ref{I_e_MG} and \ref{Nemax_vs_Nion}. Although not included in this work, the competing ionic diffusion and the reaching of a saturation of the oxygen vacancy concentration in the TaO$_{\rm x}$ layer, are expected to mitigate and eventually stop the drift dynamics. Compared to conventional Ti/HfO$_{\rm x}$ ReRAM technology, the proposed model of a TaO$_{\rm x}$/HfO$_{\rm x}$ ReRAM device shows a similar intrinsically gradual \textit{RESET}, but a fundamentally different \textit{SET}, from abrupt to exponential. The graduality of the \textit{SET} process is explained by a new physical interpretation, from a full defect depletion/accumulation of a tiny gap in the HfO$_{\rm x}$ filament to a partial modulation of defects in a larger 3D half-spherical shape in the TaO$_{\rm x}$ layer. Additionally, the narrower resistive window (typically $<10$)\cite{TEC2} in TaO$_{\rm x}$/HfO$_{\rm x}$ ReRAM devices yields a decreased temperature gradient during \textit{SET}, enhancing its graduality. \\
\textbf{The role of the CMO layer: $\sigma_{\rm CMO}$ and $k_{\rm CMO}$.} 
The proposed physical model is applicable to a range of analog bilayer ReRAM systems, including but not limited to TaO$_{\rm x}$/HfO$_{\rm x}$ ReRAM devices. Similarly to TaO$_{\rm x}$, other CMOs can be deposited in stable sub-stoichiometric phases.\cite{Falcone2023IMW} By replacing TaO$_{\rm x}$ with a designable CMO, the impact of its electrical $\sigma_{\rm CMO}$ and thermal $k_{\rm CMO}$ conductivities on the average electric field and temperature distributions is assessed in CMO/HfO$_{\rm x}$ ReRAM systems. To achieve this, the filament radius reported in Table \ref{Table2} is used, along with a constant applied bias of \textit{V} = 0.9~V on the top electrode of the CMO/HfO$_{\rm x}$ ReRAM device, assumed in LRS. Considering an arbitrary constant $k_{\rm CMO}$ = 1~W(mK)$^{-1}$,  like the one used for the TaO$_{\rm x}$ layer, Figure \ref{Fig6}a illustrates how the effectiveness of field confinement in the CMO layer (where $E_{\rm CMO}$ $\gg$ $E_{\rm CF}$) diminishes with the increase in CMO electrical conductivity $\sigma_{\rm CMO}$. Similarly, considering a fixed $\sigma_{\rm CMO}$ = $2 \cdot 10^3$~Sm$^{-1}$, like the one used for the TaO$_{\rm x}$ layer, Figure \ref{Fig6}b shows the thermal confinement in the CMO layer as a function of its thermal conductivity $k_{\rm CMO}$. For metallic metal-oxide, field confinement occurs within the conductive filament ($E_{\rm CF}$ $\gg$ $E_{\rm CMO}$) and the temperature is equally distributed between the CMO and the CF, leading to the conventional abrupt switching process dominated by the CF in Ti/HfO$_{\rm x}$ ReRAM device. On the other hand, properly designing the CMO as a thermal and electric field confinement layer is expected to yield an analog bilayer ReRAM device according to the physical model described in this work.
\begin{figure}[h]
\centering
  \includegraphics[height=4.2cm]{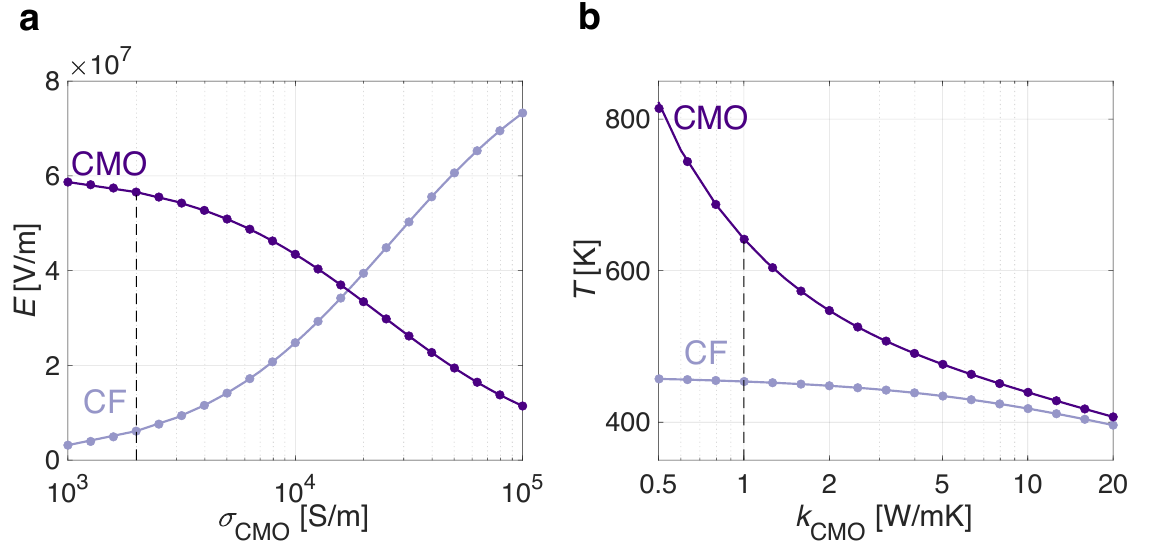}
  \caption{Considering a filamentary CMO/HfO$_{\rm x}$ ReRAM device in LRS with a constant applied bias of \textit{V} = 0.9~V, the role of the electro-thermal properties of a designable CMO material is illustrated. By increasing the CMO electrical and thermal conductivity toward values typical of a metal, the spreading resistance and the resulting effectiveness of the electric field and temperature confinement in the CMO layer, decreases. The dashed lines indicate the values used in this work for the TaO$_{\rm x}$ material.}
  \label{Fig6}
\end{figure}

%
\section*{Conclusions}
Advancing the physical understanding of the resistive switching process in analog CMO/HfO$_{\rm x}$ ReRAM devices is fundamental to further improving the hardware performance and enabling on-chip training of large neural network models with this technology. In this work, the first physics-based analytical current transport model of analog CMO/HfO$_{\rm x}$ ReRAM devices is presented. As a case study, the model is employed to simulate experimental data from a representative analog TaO$_{\rm x}$/HfO$_{\rm x}$ ReRAM device. Table \ref{Table3} presents a comparison of this work with available ReRAM models in literature. Based on this work, future studies will focus on the development of a compact model description including device failure mechanisms, and subsequently circuit simulations of analog CMO/HfO$_{\rm x}$ ReRAM technology.
\begin{table}[h]
\small
  \caption{\ Comparison with available ReRAM models in literature}
  \label{Table3}
  \begin{tabular*}{0.48\textwidth}{@{\extracolsep{\fill}}lllll}
    \hline
     & Bengel et.al. \cite{Bengel2020} & Wu & Falcone & This \\
     & Jiang et.al. \cite{Stanford2016} & et. al.\cite{Wu2017_TaOxTEL} & et. al. \cite{Falcone2023IMW}& work\\
    \hline
    Device & ReRAM & ReRAM & ReRAM & ReRAM \\
    Materials & HfO$_{\rm x}$ & TaO$_{\rm x}$/HfO$_{\rm x}$ & CMO/HfO$_{\rm x}$ & TaO$_{\rm x}$/HfO$_{\rm x}$ \\
    Synapse & Digital & Analog & Analog & Analog \\
    Model & Analytical & Empirical & Empirical & Analytical \\
    Transport & $e$ and $V_{\rm O}^{\rm \cdot \cdot}$ & $e$ & $e$ & $e$ and $V_{\rm O}^{\rm \cdot \cdot}$ \\
    Approx. & Compact model  & FEM & FEM & FEM \\
    \hline
  \end{tabular*}
\end{table} \\
In this work, it is shown that the combination of the filamentary current constriction with the electro-thermal properties of the TaO$_{\rm x}$ material, enables temperature and electric field confinement in the TaO$_{\rm x}$ layer. Consequently, the resistive switching is dominated by the defect dynamics in the TaO$_{\rm x}$ layer rather than in the conductive filament as per conventional M/I/M ReRAM stack. The proposed analytical current transport model describes the Mott-Gurney trap-to-trap tunneling process. The resistive switching of the analog ReRAM device is attributed to the modulation of the density of defect states in the sub-band of the TaO$_{\rm x}$ layer. During the \textit{RESET} switching process, oxygen vacancies in the TaO$_{\rm x}$ layer radially migrate toward the conductive filament, partially depleting a half-spherical volume, also known as dome. As a consequence, the lower density of defect states results in a lower electrical conductivity of the TaO$_{\rm x,Dome}$. In this process, the electric field increases while the temperature decreases. Conversely, the \textit{SET} process, characterized by an exponentially increasing temperature and a nearly constant electric field, consists of a defect-enrichment of the TaO$_{\rm x,Dome}$, restoring the LRS condition. The more gradual \textit{SET} of TaO$_{\rm x}$/HfO$_{\rm x}$ ReRAM devices compared to conventional Ti/HfO$_{\rm x}$ technology is attributed to the change from a full defect depletion/accumulation of a tiny gap in the conductive filament to a partial defect modulation in a larger TaO$_{\rm x,Dome}$. Additionally, the narrower resistive window in TaO$_{\rm x}$/HfO$_{\rm x}$ ReRAM device yields a decreased temperature gradient during the \textit{SET}, enhancing its graduality. The intermediate resistive states are described as a gradual increase/decrease of the TaO$_{\rm x,Dome}$ defect density, which results in a variation of its electrical conductivity. The \textit{SET} and \textit{RESET} activation energies, extracted from the modelling of the oxygen vacancy drift dynamics, closely align with the reported oxygen vacancy migration energy in the TaO$_{\rm x}$ layer. The physics-based analytical current transport model developed in this work is not limited to TaO$_{\rm x}$/HfO$_{\rm x}$ devices, but applies to other similar analog CMO/HfO$_{\rm x}$ ReRAM systems. Eventually, the role of the $\sigma_{\rm CMO}$ and $k_{\rm CMO}$ on the electric field and temperature distributions is described, setting the groundwork for further CMO/HfO$_{\rm x}$ ReRAM device optimization.
\section*{Author Contributions}
Conceptualization: D.F.F. and V.B.; Methodology and validation: D.F.F., S.M. and V.B.; Simulations: D.F.F.; Result interpretation: D.F.F., S.M., M.G., T.S., V.B., A.L.P and B.J.O.; Data curation: D.F.F. and V.B.; Hardware fabrication: T.S.; Electrical characterization: D.F.F. and T.S.; Supervision: V.B. and S.M; Manuscript writing: D.F.F.; Manuscript review and editing: All authors; Funding acquisition: B.J.O. and V.B.

\section*{Conflicts of interest}
There are no conflicts to declare.

\section*{Acknowledgements}
The authors acknowledge the Binnig and Rohrer Nanotechnology Center (BRNC) at IBM Research Europe - Zurich. This work is funded by SNSF \textit{ALMOND} (grantID: 198612), by the European Union and Swiss state secretariat SERI within the H2020 \textit{MeM-Scales} (grantID: 871371), \textit{MANIC} (grantID: 861153) and \textit{PHASTRAC} (grantID: 101092096) projects and by the Federal Ministry of Education and Research (BMBF, Germany) within the \textit{NEUROTEC} (grantID: 16ME0398K) project.







\scriptsize{
\providecommand*{\mcitethebibliography}{\thebibliography}
\csname @ifundefined\endcsname{endmcitethebibliography}
{\let\endmcitethebibliography\endthebibliography}{}

}

\end{document}